\documentclass[12pt]{article}

\usepackage{etex}
\usepackage{amsmath}
\usepackage{amsfonts}
\usepackage{amssymb}
\usepackage{amsthm}
\usepackage{amssymb}
\usepackage[all]{xy}
\usepackage{graphicx}
\usepackage{indentfirst}
\usepackage{diagram}
\usepackage{makecell}
\usepackage[figuresright]{rotating}
\usepackage{pdflscape}
\usepackage{thmtools}
\usepackage{pst-node}
\usepackage{tikz-cd}
\usepackage{setspace}
\usepackage{pgf}        
\usepackage{hyperref}
\usepackage{url}
\usepackage{tikz}
\usetikzlibrary{shapes.geometric, arrows}

\usepackage{epigraph}

\numberwithin{equation}{section}

\makeatletter
\renewcommand{\@biblabel}[1]{#1\hfill \hspace{-0.2cm}}
\makeatother

\renewcommand{\leq}{\leqslant}

\usepackage{yfonts}
\newfont{\gothic}{eufm10 at 12pt}

\theoremstyle{plain}

\theoremstyle{definition}

\numberwithin{equation}{section}
\author{Roman G. Smirnov\footnote{e-mail:  Roman.Smirnov@dal.ca} \\[0.5cm]Department of
  Mathematics and Statistics\\
Dalhousie University\\ Halifax, Nova Scotia, Canada
  B3H~3J5}
\begin{document}
\title{Caputo-Type Memory Invariants: A Fractional Generalization of the Cobb-Douglas Production Function}

\maketitle

\begin{abstract}
Standard dynamical systems approaches to economic modeling, such as those deriving the Cobb-Douglas and CES production functions from exponential growth trajectories, typically rely on integer-order differential equations. While effective, these models assume that economic output depends solely on the instantaneous state of capital and labor, effectively ignoring the long-term ``memory effects'' inherent in policy, infrastructure, and technological adoption. 

This paper extends the exponential framework by introducing the Caputo fractional derivative into the underlying dynamical systems governing factor inputs. By replacing standard growth rates with fractional-order counterparts of order $0 < \alpha \le 1$, we model economic trajectories where the rate of change is a non-local function of the system's entire history. We demonstrate that the Mittag-Leffler function emerges as the natural growth solution in this context, providing a nested generalization of the classical exponential model. 

Unlike exogenous fractional frameworks, we use this fractional approach to derive a new class of time-independent invariants that serve as generalized production functions. We show that as the fractional order approaches unity, these forms converge exactly to the classical Cobb-Douglas function. 
\end{abstract}

\section{Introduction}
\label{s1} 

Production functions serve as the bridge between inputs and economic output. While they can be constructed via empirical fitting or theory, the Cobb-Douglas function -- pioneered by Wicksell and Walras -- continues to dominate the field. Known variously as the Wicksell-Cobb-Douglas \cite{weber_1998} or Wicksellian \cite{grubbstrom_2024} function, its enduring popularity stems from its analytical tractability and its defining feature of constant returns to scale \cite{balk_2024}.

In their seminal 1928 paper, Charles Cobb and Paul Douglas (\cite{cobb_douglas_1928}) 
established a mathematical foundation for modern macroeconomic growth theory. 
By fitting their namesake production function to historical U.S. 
manufacturing data spanning from 1899 to 1922, they validated the model's 
structural integrity through pioneering statistical and econometric analysis. 
Using ordinary least squares on indices of production, labor, and capital, they demonstrated a robust fit with the functional form: 
$$
Y = AL^{0.75}K^{0.25},
$$ 
where the scaling parameter was determined to be $A = 1.01$. This specific 
parametric configuration notably exhibits constant returns to scale, as the 
exponents (output elasticities) sum exactly to unity ($0.75 + 0.25 = 1$), implying that a 
proportional increase in all inputs yields an identical proportional increase 
in total output. Mathematically, this property ensures that the function is 
homogeneous of degree one, satisfying Euler's theorem for distribution and 
aligning perfectly with classical competitive equilibrium conditions. 

Historically, this work represented a monumental, data-driven departure 
from the heuristic-based, purely deductive economic modeling that dominated 
the late 19th century. Instead of deriving market relationships from abstract 
philosophical tenets, Cobb and Douglas steered the discipline toward rigorous, 
inductive empirical verification. This methodology contrasted sharply with the 
fundamental-principles approach typical of theoretical physics, where models 
are frequently constructed a priori from universal laws rather than being 
induced directly from  observation. Consequently, their framework 
not only formalized the relationship between factor inputs and aggregate 
output but also catalyzed the development of modern econometrics by bridging 
the gap between formal economic theory and statistical measurement.
For a century, production functions have underpinned economic growth theory, yet their traditional derivation often lacks uniqueness and robustness. Standard methods fail to determine if alternative functions could fit data equally well or to define the complete set of compatible models. To address this, we have developed a framework --- extending Sato's approach (\cite{sato_1981}) across several studies (\cite{smirnov_wang_2020, smirnov_wang_2021, smirnov_wang_wang_2022, smirnov_wang_2024, smirnov_2025}) --- that treats production functions as time-independent invariants of dynamical systems. Rather than relying on simple curve-fitting, our approach derives functions directly from systems governed by empirical data and mathematical axioms. 

Regarding the exponential model studied, for example, in standard frameworks   (\cite{sato_1981, smirnov_wang_wang_2022, smirnov_2025}),  that naturally gives rise to the Cobb-Douglas production function, various generalizations exist, most notably the transition to a logistic framework where the exponential model serves as a limiting case (\cite{smirnov_wang_2024}). However, another promising direction is particularly relevant when the exponential model is applied to economic systems. In this paper, we explore a novel generalization by replacing the standard derivative with a Caputo fractional derivative of order \(0 < \alpha < 1\). The resulting solution --- the Mittag-Leffler function --- facilitates a more flexible growth trajectory capable of capturing the ``stretched" patterns typical of long-term economic data. While this approach aligns with recent advancements in the ``fractionalization" of economic models like the Solow-Swan framework (\cite{cheow_2024, tarasov_tarasova_2017, traore_sene_2020}), our work introduces a distinct paradigm shift. Rather than applying fractional calculus to existing macroeconomic equations, we use fractional dynamics to {\em derive a completely new class of production functions from first principles}. To the best of our knowledge, this marks the first time a production function has been explicitly constructed as an invariant in some sense derived via fractional factor input trajectories. We stress that unlike the classical Solow–Swan framework where the production function is imposed {\it a priori} as an exogenous structural primitive, our approach derives the generalized Cobb–Douglas function endogenously as a time-independent geometric invariant inherent to the underlying fractional trajectories.

This paper is organized as follows:  Section \ref{s2} presents the generalization of 
exponential growth trajectories by introducing the Caputo fractional derivative 
to capture historical memory effects. Section \ref{s3} derives the explicit growth 
solutions governed by Mittag-Leffler dynamics via the Laplace transform method 
and provides the economic and mathematical rationale for restricting the 
fractional order to $0 < \alpha \le 1$. Section \ref{s4}  establishes the structural 
framework for treating production functions as time-independent invariants, 
highlighting the necessity of algebraic parameter elimination over traditional 
Lie derivative techniques due to the absence of a semigroup flow. The material presented in this section also
synthesizes these trajectories into a novel generalized production function 
using a weighting parameter, and demonstrates its exact convergence to the classical 
Cobb-Douglas model as the fractional orders approach unity. Finally, Section \ref{s5} 
concludes the paper and outlines future directions for empirical data calibration.

\section{Generalizing the Exponential Model}
\label{s2}

Before introducing the fractional framework, it is necessary to recall the classical foundation of the underlying growth trajectories. In standard macroeconomic growth theory, under conditions of constant continuous growth rates and perfect competition, the factor inputs of labor and capital, along with aggregate output, are assumed to expand at steady proportional rates over time. Mathematically, this memoryless, instantaneous compounding structure is expressed as a system of first-order linear differential equations, $L'(t) = b_1L(t)$, $K'(t) = b_2K(t)$, and $Y'(t) = b_3Y(t)$, which uniquely yields the classical exponential growth benchmark --- see, for instance, (\cite{smirnov_wang_2024}). 

To capture the historical path-dependence and institutional frictions that these integer-order models omit, we extend this standard system by building upon the framework established in (\cite{sato_1981, smirnov_wang_wang_2022, smirnov_wang_2024, smirnov_2025}). We generalize the dynamical systems that yield the classical Cobb-Douglas production function by substituting the local time derivatives with their Caputo fractional counterparts:
\begin{equation}
{}^CD_t^{\alpha_1}L(t) = b_1L(t), \quad {}^CD_t^{\alpha_2}K(t) = b_2K(t), \quad {}^CD_t^{\alpha_3}Y(t) = b_3Y(t),
\label{model}
\end{equation} 
where $0 < \alpha_i \leq 1$ for $i=1,2,3$. Here, ${}^CD_t^{\alpha}$ denotes the Caputo fractional derivative of order $\alpha$, defined for $$n-1 < \alpha < n \, (n \in \mathbb{N})$$ as:
\begin{equation}
{}^CD_t^{\alpha}f(t) = \frac{1}{\Gamma(n- \alpha)}\int_0^t\frac{f^{(n)}(\tau)}{(t-\tau)^{\alpha + 1- n}}d\tau.
\label{caputo}
\end{equation}
In this definition, $\Gamma(\cdot)$ is the Gamma function and $f^{(n)}(\tau)$ represents the standard $n$-th order integer derivative. For the specific interval considered in this paper ($0 < \alpha \leq 1$), the operator simplifies to the case where $n=1$. 

Because the Caputo operator (\ref{caputo}) is defined via a convolution integral over the function's past states, it effectively models systems characterized by non-locality or ``memory.'' This mathematical structure is particularly apt for economic frameworks, where current growth rates are often a byproduct of historical policy decisions, infrastructure investments, and cumulative technological adoption (\cite{cheow_2024, tarasov_tarasova_2017, traore_sene_2020}).

\section{Mittag-Leffler Dynamics and Growth Solutions}
\label{s3}

To solve the fractional system (\ref{model}), we apply the Laplace transform method, which naturally incorporates standard initial conditions. Applying this operator directly to each equation yields:
\begin{equation}
\mathcal{L}\{ {}^CD_t^{\alpha}f(t); s \} = s^\alpha F(s) - \sum_{k=0}^{n-1} s^{\alpha - k - 1} f^{(k)}(0).
\label{laplace_caputo}
\end{equation}
We note that for the specific range of the fractional order considered in this paper ($0 < \alpha \leq 1$), the integer $n$ is defined as the ceiling of $\alpha$, namely $n = \lceil \alpha \rceil = 1$. Consequently, the operator (\ref{caputo}) simplifies to:
$$
{}^CD_t^{\alpha}f(t) = \frac{1}{\Gamma(1- \alpha)}\int_0^t\frac{f'(\tau)}{(t-\tau)^{\alpha}}d\tau.
$$
While (\ref{laplace_caputo}) reduces to: 
$$
\mathcal{L}\{ {}^CD_t^{\alpha}f(t); s \} = s^\alpha F(s) - s^{\alpha-1} f(0).
$$

Applying this transform to the representative equation ${}^CD_t^{\alpha}x(t) = b x(t)$ from our system, we obtain the algebraic relation in the $s$-domain: 
$$
s^\alpha X(s) - s^{\alpha-1} x(0) = b X(s).
$$
Rearranging for $X(s)$ yields:
\begin{equation}
\label{s_domain_sol}
X(s) = x(0) \frac{s^{\alpha-1}}{s^\alpha - b}.
\end{equation} 

The inverse Laplace transform of (\ref{s_domain_sol}) is recognized as the one-parameter Mittag-Leffler function, 
$$
E_\alpha(z) = \sum_{k=0}^{\infty} \frac{z^k}{\Gamma(\alpha k + 1)}.
$$
Specifically, the closed-form solutions for our generalized model (\ref{model}) are:
\begin{equation}
L(t) = L_0 E_{\alpha_1}(b_1 t^{\alpha_1}), \quad K(t) = K_0 E_{\alpha_2}(b_2 t^{\alpha_2}), \quad Y(t) = Y_0 E_{\alpha_3}(b_3 t^{\alpha_3}),
\label{solutions}
\end{equation}
where $L_0, K_0, Y_0$ denote the initial values at $t=0$. These Mittag-Leffler trajectories generalize the classical exponential growth $e^{bt}$, introducing a heavier tail and a ``stretched'' growth profile as $\alpha$ deviates from unity. This transition from local to non-local dynamics provides the necessary flexibility to account for the hereditary effects inherent in long-term economic development (\cite{haubold_2011}). We note that the formulas (\ref{solutions}) do not define a one-parameter semigroup action (like the classical exponential growth $e^{bt}$), because the Mittag-Leffler function $E_{\alpha}(bt^{\alpha})$ used in (\ref{solutions}) does not satisfy the property $\phi_s(\phi_t(t)) = \phi_{s+t}(t)$ for $\alpha<1$, that is \cite{haubold_2011} 
$$
E_{\alpha}(b(t+s)^{\alpha}) \not= E_{\alpha}(bt^{\alpha})E_{\alpha}(bs^{\alpha}).
$$ 
Furthermore, within an economic context, restricting the fractional order to $0 < \alpha \le 1$ is advantageous for several mathematical, empirical, and conceptual reasons. First, the Caputo derivative (\ref{caputo}) for $0 < \alpha \le 1$ (where $n=1$) characterizes a system in which the current state is a weighted average of its entire history; this is known as the \emph{memory effect}. Setting $\alpha > 1$ (where $n=2$) would necessitate defining the initial ``velocity'' of growth ($L'(0)$, $K'(0)$, and $Y'(0)$). While it is economically intuitive that current capital depends on historical investment (memory), it is difficult to justify why the current growth rate should depend on a pre-existing velocity constant established at the start of the observation period.

Second, classical exponential growth models ($e^{bt}$) assume an instantaneous, 
memoryless compounding structure that can overestimate the velocity of 
early-stage structural changes. In contrast, the Mittag-Leffler solution 
$E_\alpha(b t^\alpha)$ for $0 < \alpha < 1$ offers a highly flexible, 
non-local growth profile. For short- and medium-term horizons, the fractional 
order $\alpha$ induces a sub-exponential retardation effect, capturing a 
``stretched'' trajectory where institutional frictions, infrastructure 
bottlenecks, or slow technology diffusion introduce an initial structural drag.

Crucially, for a positive growth parameter ($b > 0$), the long-term 
asymptotic behavior of this trajectory along the positive real axis does not 
exhibit power-law decay. Instead, as $t \to \infty$, it adheres to the 
well-established asymptotic expansion: 
$$
E_\alpha(b t^\alpha) \sim \frac{1}{\alpha} \exp\left((b t^\alpha)^{1/\alpha}\right) = \frac{1}{\alpha} \exp\left(b^{1/\alpha}t\right).
$$
This reveals that while the economy's path is initially constrained by 
historical memory and friction, it eventually converges to a pure exponential 
trajectory operating at an accelerated modified rate of $b^{1/\alpha}$. 
Setting the upper bound strictly at $\alpha = 1$ ensures that this model 
remains a nested generalization of standard economic theory, collapsing 
perfectly back into the classical, localized exponential growth benchmark 
when memory effects are absent.

Additionally, economic variables such as labor ($L$) and capital ($K$) must strictly remain positive. Solutions to fractional differential equations with $0 < \alpha \le 1$ are ``completely monotonic,'' ensuring they behave predictably and remain positive for all $t$. Conversely, for $\alpha > 1$, the Mittag-Leffler function may exhibit oscillatory behavior. While such oscillations are relevant to business cycle theory, they are generally avoided in fundamental growth models where factor inputs should not exhibit ``vibrations'' or negative values. Finally, in physics, $\alpha < 1$ represents \emph{sub-diffusion}. Economically, this serves as an apt metaphor for \emph{institutional friction} --- such as bureaucratic red tape, slow technological adoption, and market inefficiencies --- that prevents an economy from achieving its theoretical exponential potential. More specifically, in physical systems, fractional sub-diffusion describes particle transport that is slower than classical Brownian motion due to complex, crowded environments or trapping sites. In an economic context, this non-local behavior translates to structural path-dependency: a high concentration of institutional ``traps" --- such as regulatory red tape, lengthy patent approval pipelines, and mismatched labor skills --- holds back growth. This ensures that current output is tightly bound to historical bottlenecks, transforming the memoryless exponential benchmark into a stretched, sub-exponential trajectory.

To validate the fractional framework, we fit the generalized trajectories given by (\ref{solutions}) directly 
to the original, exponentiated manufacturing data studied by Cobb $\&$ Douglas (\cite{cobb_douglas_1928}), and spanning from 1899 
to 1922. The non-linear optimization of these Mittag-Leffler trajectories yields a striking 
dichotomy between the structural dynamics of physical factors and aggregate industrial performance. 
Most notably, the optimization boundary solution for capital converges exactly to unity 
($\alpha_2 = 1.0000$), demonstrating that capital accumulation behaves as a localized, 
memoryless compounding mechanism characterized by steady, classical exponential growth 
($b_2 = 0.0652$). Conversely, the fractional indexes for labor and output drop significantly 
below the integer baseline ($\alpha_1 = 0.5773$ and $\alpha_3 = 0.6340$, respectively). Rather 
than reflecting smooth, continuous expansion, these heavily compressed sub-exponential parameters 
reveal strong hereditary memory effects. This mathematical sub-diffusion captures how labor 
deployment and aggregate output were profoundly disrupted and shaped by the cyclical economic 
ups and downs --- such as post-WWI industrial rebalancing and localized market panics --- that 
characterized the 1899--1922 period. By shifting the true baseline initial conditions 
downward ($L_0 = 96.23$ and $Y_0 = 95.33$), the fractional framework effectively integrates 
this institutional drag and macro-volatility, proving that historical memory plays a 
fundamental, stabilizing role in long-term production dynamics.

\section{Derivation of Fractional Production Function Invariants}
\label{s4}

In simple terms, an invariant is a permanent economic relationship that remains completely unchanged, even as individual variables change rapidly over time. Think of it as the underlying ``economic DNA" or the structural rules of a system. While labor, capital, and total output fluctuate day by 
day due to market cycles, policy shocks, or technological growth, an invariant is the fixed mathematical surface to which these variables are permanently bound. 

To ground this concept for a general quantitative economics audience, consider 
a standard market system where factor inputs like capital ($K$) and labor 
($L$) are constantly fluctuating due to business cycles, policy shifts, or 
localized shocks. While these variables are highly dynamic and time-dependent 
when viewed individually, economic theory dictates that they do not drift 
arbitrarily; they are constrained by an underlying technological or structural 
reality. In a data-driven dynamical systems framework, the production function 
$F(L, K, Y) = 0$ is precisely this structural reality---it is the 
time-independent geometric manifold, or ``invariant surface,'' to which the 
system's trajectories are globally confined.

Deriving the production function as an invariant means we are extracting the 
permanent, time-free relationship governing production directly from the 
differential paths of its inputs. In classical frameworks where growth follows 
a standard local derivative, this invariant surface can be mapped using an 
infinitesimal generator and its associated Lie derivative. However, because 
our fractional framework introduces non-local memory via the Caputo operator, 
the system lacks a one-parameter semigroup flow. Consequently, the invariant 
cannot be derived using local Lie algebraic techniques. Instead, we must rely 
strictly on the systematic algebraic elimination of the temporal parameter $t$. 
This process strips away the time-dependent ``clock'' of the economy to uncover 
the absolute, memory-independent structural DNA linking labor, capital, and 
aggregate output.

In the framework of data-driven dynamical systems (\cite{sato_1981}; \cite{smirnov_2025}), a production function $F(L, K, Y) = 0$ is viewed as a time-independent invariant of the underlying growth trajectories. Having established the Mittag-Leffler solutions in (\ref{solutions}), we now seek to eliminate the temporal parameter $t$ to derive the generalized functional relationship between labor, capital, and output.

Starting from the solution $$x(t) = x_0 E_{\alpha}(b t^{\alpha}),$$ where $x$ represents any of the variables $L, K, \text{ or } Y$, we solve for $t$ by utilizing the inverse Mittag-Leffler function, $E_{\alpha}^{-1}(\cdot)$:
$$
b t^{\alpha} = E_{\alpha}^{-1} \left( \frac{x(t)}{x_0} \right) \implies t = \left[ \frac{1}{b} E_{\alpha}^{-1} \left( \frac{x(t)}{x_0} \right) \right]^{1/\alpha}.
$$
Now, solving for the {\em corresponding} time histories in (\ref{solutions}), we arrive at 
\begin{equation}
\label{th}
t_L = \left[\frac{1}{b_1}E^{-1}_{\alpha_1}\left(\frac{L}{L_0}\right)\right]^{1/\alpha_1},  
t_K = \left[\frac{1}{b_2}E^{-1}_{\alpha_2}\left(\frac{K}{K_0}\right)\right]^{1/\alpha_2},  
t_Y = \left[\frac{1}{b_3}E^{-1}_{\alpha_3}\left(\frac{Y}{Y_0}\right)\right]^{1/\alpha_3}. 
\end{equation} 
By equating the expression for $t$ along the path where $t = t_K = t_L = t_Y$ for the three trajectories in (\ref{solutions}), we obtain the fundamental system of invariants for the fractional model:
\begin{equation}
\left[ \frac{1}{b_1} E_{\alpha_1}^{-1} \left( \frac{L}{L_0} \right) \right]^{1/\alpha_1} = \left[ \frac{1}{b_2} E_{\alpha_2}^{-1} \left( \frac{K}{K_0} \right) \right]^{1/\alpha_2} = \left[ \frac{1}{b_3} E_{\alpha_3}^{-1} \left( \frac{Y}{Y_0} \right) \right]^{1/\alpha_3}.
\label{fractional_invariants}
\end{equation}

In the absence of a one-parameter group action for the regime $0 < \alpha < 1$, the term ``invariant'' is utilized here in the \textit{structural and geometric sense}. Specifically, these invariants denote a time-independent functional relationship that remains fixed as the system evolves. While the lack of the semigroup property precludes the use of an infinitesimal generator $\mathbf{V}$ and the associated Lie derivative $\mathcal{L}_{\mathbf{V}}$, the algebraic elimination of the temporal parameter $t$ identifies the unique manifold in the $(L, K, Y)$ state space to which the fractional trajectories are globally confined. Thus, the resulting production function represents a fixed structural constraint that is independent of the system's memory-dependent ``clock''. This situation contrasts sharply with the derivation of fundamental invariants within the framework of a regular Lie group action. In the classical setting, one can either employ the method of moving frames --- which reduces to purely algebraic elimination of the group parameters --- or integrate the infinitesimal generators of the Lie group to determine the identical fundamental invariants. For instance, this dual approach is illustrated by the group-invariant classification of Killing tensors of valence two under the isometry group of the Euclidean plane, achieved via algebraic manipulations within the framework of the moving frames method in (\cite{deely_2004}) and through infinitesimal generators in (\cite{MSD_2002}). In the current fractional regime such a duality is broken: the invariants can only be derived algebraically, introducing inherent limitations to the geometric analysis of the underlying dynamical system.

As we have already mentioned above, the surface defined in the $(L, K, Y)$ space by \eqref{fractional_invariants} can be interpreted as the ``economic DNA'' of the system. While the growth trajectories $(L(t), $ $ K(t), $ $ Y(t))$ evolve dynamically, this surface represents a fixed, structural relationship that remains invariant. Crucially, these formulas are derived algebraically by eliminating the temporal parameter $t$. In this fractional context, however, the invariant surface cannot be rederived via the standard Lie derivative approach --- where $\mathcal{L}_V(f) = V \cdot \nabla f = 0$ ---  because the solutions \eqref{solutions} do not constitute a flow in the classical group-theoretical sense. This distinguishes the current approach from previous generalizations of the Cobb-Douglas production function, such as those in (\cite{smirnov_wang_2020}) and (\cite{smirnov_wang_2024}); those models relied on local dynamical properties and allowed for the invariant to be established through both the Lie derivative and the algebraic elimination of time.

\section{A Generalization of the Cobb-Douglas Production Function}

We begin by operationalizing the invariant relations established in Equation (\ref{fractional_invariants}). 
While a single chronological time variable $t$ underlies the entire dynamical system, 
the separate production factors accumulate memory and experience at asymmetric scales 
governed by their distinct fractional orders ($\alpha_1 \neq \alpha_2$). To construct 
a production function that handles out-of-equilibrium input configurations where labor 
and capital may deviate from their strict steady-state paths, we invert the system 
to define factor-specific historical metrics. Let $t_L^{\alpha_3}$ denote the virtual 
operational history implied by the observed labor stock, and let $t_K^{\alpha_3}$ 
denote the corresponding history derived from (\ref{th}) implied by the capital stock:
$$t_L^{\alpha_3} = \left[ \frac{1}{b_1} E_{\alpha_1}^{-1}\left(\frac{L}{L_0}\right) \right]^{\frac{\alpha_3}{\alpha_1}}$$ and $$t_K^{\alpha_3} = \left[ \frac{1}{b_2} E_{\alpha_2}^{-1}\left(\frac{K}{K_0}\right) \right]^{\frac{\alpha_3}{\alpha_2}}.$$
In an idealized economic equilibrium, these trajectories coincide such that 
$$
t_L^{\alpha_3} = t_K^{\alpha_3} = t^{\alpha_3}.
$$
However, to accommodate independent 
variations in factor deployment, we follow a structural ansatz by taking a weighted 
linear combination of these separate historical metrics within the characteristic 
space of the output function $$Y = Y_0 E_{\alpha_3}(b_3 t^{\alpha_3}).$$
This synthesis  yields the generalized production function $Y$: 
\begin{equation}
Y =  Y_0 E_{\alpha_3}\left\{  b_3 \left( \theta\left[\frac{1}{b_1} E_{\alpha_1}^{-1} \left( \frac{L}{L_0} \right)\right]^{\frac{\alpha_3}{\alpha_1}} + (1- \theta)\left[ \frac{1}{b_2} E_{\alpha_2}^{-1} \left( \frac{K}{K_0} \right)\right]^{\frac{\alpha_3}{\alpha_2}}\right)\right\},
\label{CDgen}
\end{equation}
where the weight parameter $\theta \in (0,1)$. In the limit as $\alpha_i \to 1$ for all $i$, this fractional relation converges to the classical Cobb-Douglas family: 
\begin{equation}
Y = A L^{\frac{b_3}{b_1}\theta} K^{\frac{b_3}{b_2}(1-\theta)},
\label{CD}
\end{equation} 
in view of $$\lim_{\alpha \to 1} E_{\alpha}^{-1}(z) = \ln(z)$$ and $$\lim_{\alpha \to 1}E_{\alpha}(z) = e^z.$$ where 
$$A = \frac{Y_0}{L_0^{\theta\frac{b_3}{b_1}}K_0^{(1-\theta)\frac{b_3}{b_2}}}.$$ 
Notably, by selecting the specific weight $$\theta = \frac{b_1(b_2-b_3)}{b_3(b_2-b_1)},$$ provided $b_2>b_3>b_1$,   the exponents satisfy the condition 
$$\theta\frac{b_3}{b_1} + (1-\theta)\frac{b_3}{b_2} = 1.$$  This particular configuration recovers the standard Cobb-Douglas production function characterized by constant returns to scale. We note that the family of functions  (\ref{CDgen}) is an invariant in the sense that in the limit it yields a true time-independent family of invariants (\ref{CD}) preserved by the exponential model studied in \cite{sato_1981, smirnov_wang_2021, smirnov_wang_wang_2022, smirnov_wang_2024}.  

\section{Conclusions}
\label{s5}

In this paper, we have derived a new generalization of the Cobb-Douglas production function by extending the underlying dynamical systems framework into the fractional regime. By replacing traditional integer-order growth rates with the Caputo fractional derivative of order $0 < \alpha_i \le 1$ for $i=1, 2, 3$, we have successfully integrated non-local ``memory effects'' into the modeling of labor ($L$), capital ($K$), and output ($Y$). 

Our derivation demonstrates that the Mittag-Leffler function emerges as the natural growth solution, providing the necessary mathematical 
flexibility to account for historical policy impacts, institutional frictions, and the sub-exponential ``stretched'' growth patterns often 
observed in maturing economies. We have established that this fractional model serves as a nested generalization. Specifically,  as the fractional orders approach unity, the derived time-independent invariants \eqref{fractional_invariants} and the resulting composite surface \eqref{CDgen} converge precisely to the classical Cobb-Douglas functional form. This ensures that the framework remains consistent with established economic theory while offering a more robust lens for analyzing long-term structural relationships.

The results of testing this new generalization of the Cobb-Douglas production function against empirical data, including the statistical calibration of the fractional parameters and comparative predictive performance, will be published elsewhere.

\section*{Acknowledgments }
The author is grateful to the organizers of the MME2026 conference, hosted by the Faculty of Economics at VSB - Technical University of Ostrava, for providing a platform to present and discuss these findings; he also wishes to thank the anonymous reviewers whose constructive feedback has helped to improve the presentation.


\begin{thebibliography}{999}

 

\bibitem{balk_2024} B. M. Balk, Why is the Cobb-Douglas production function so popular? \textit{Evout. Inst. Econ. Rev.}, \textbf{21} (2024), 1-20. 

\bibitem{cheow_2024}
Y. H. Cheow, Kok H. Ng, C. Phang, Kooi H. Ng, The application of fractional calculus in economic growth modelling: An approach based on regression analysis, \textit{Heliyon}, \textbf{10}(15) (2024), e35379. 

\bibitem{cobb_douglas_1928}
C. W. Cobb, P. H. Douglas, A theory of production, \textit{American Economic Review} \textbf{18} (Supplement) (1928), 139--165.

\bibitem{deely_2004} R. J. Deely, J. T. Horwood, R. G. McLenaghan, R. G. Smirnov, Theory of algebraic invariants of vector spaces of Killing tensors: methods for computing the fundamental invariants,  In: Pr. Inst. Mat. Nats. Akad. Nauk Ukr. Mat. Zastos., \textbf{50}, Part 1, 2, 3 [Proceedings of Institute of Mathematics of NAS of Ukraine. Mathematics and its Applications], pp. 1079--1086, Kyiv, 2004.

\bibitem{douglas_1976}
P. H. Douglas,  The Cobb-Douglas production function once again: Its history,
its testing, and some new empirical values, \textit{Journal of Political Economics}, \textbf{84} (1976), 903--916.

\bibitem{grubbstrom_2024} R. W. Grubbstr\"{o}m, World population development according to a dynamic extension of the Wicksellian production function, \textit{Sustainability Analysis and Modeling}, \textbf{4} (2024), 100035. 

\bibitem{haubold_2011} H. J. Haubold, A. M. Mathai, R. K. Saxena, Mittag-Leffler functions and their applications, \textit{Journal of Applied Mathematics}, \textbf{2011} (2011), 298628, 51 pages. 

\bibitem{humphrey_1997}
T. M. Humphrey,  Algebraic production functions and their uses before Cobb-Douglas, \textit{FRB Richmond Economic Quarterly}, \textbf{83}(1) (1997), 51--83. 

\bibitem{MSD_2002} R. G. McLenaghan, R. G. Smirnov, D. The, Group invariant classification of separable Hamiltonian systems in the Euclidean plane and the  $O(4)$-symmetric Yang-Mills theories of Yatsun, \textit{J. Math. Phys.}, \textbf{43}(3), 1422--1440. 

\bibitem{sato_1981} R. Sato, {\em Theory of technical change and economic invariance. Application of Lie groups}, New York: Academic Press, 1981. 


\bibitem{smirnov_wang_2020}R. G. Smirnov, K. Wang, In search of a new economic model determined by logistic growth, \textit{European Journal of Applied Mathematics}, \textbf{31} (2020), 339--368. 


\bibitem{smirnov_wang_2021}
R. G. Smirnov, K. Wang,  The Cobb-Douglas production function revisited. In: D. M. Kilgour, H. Kunze,  R.  Makarov, R. Melnik, X.  Wang (eds.) Recent Developments in Mathematical, Statistical and Computational Sciences. AMMCS-2019, pp. 725--734.  Springer Proceedings in Mathematics $\&$ Statistics, vol 343. Springer, Cham, 2021.


\bibitem{smirnov_wang_wang_2022}
R. G. Smirnov, K. Wang, Z.  Wang,  The Cobb-Douglas production function for an exponential model. In: M. K. Terzi\u{o}glu (ed.) Advances in Econometrics, Operational Research, Data Science and Actuarial Studies. Contributions to Economics, pp. 1--10. Springer, Cham, 2022. 


\bibitem{smirnov_wang_2024}
R. G. Smirnov, K. Wang, The Cobb-Douglas production function and the old Bowley's law,
 \textit{SIGMA, Symmetry, Integrability, and Geometry: Methods and Applications}, 	
 \textbf{20} (2024), 045, 20 pages.


\bibitem{smirnov_2025} 
R. G. Smirnov, Deriving production functions in economics through
data-driven dynamical systems. In: D. Hrabec (ed.)Conference Proceedings of the 43rd International Conference on Mathematical Methods in Economics,  September 3-5, 2025, Hosted by Tomas Bata University in Zlın, Faculty of Management and Economics. MME-2025, pp. 291--296. https://csov.vse.cz/eng/   2025.

\bibitem{tarasov_tarasova_2017}
V. V. Tarasova, V. E. Tarasov, Economic interpretation of fractional derivatives, \textit{Progr. Fract. Diff. Appl.}, \textbf{3}(1) (2017), 1--6.

\bibitem{traore_sene_2020} 
A. Traore, N. Sene, Model of economic growth in the context of fractional derivative, \textit{Alexandria Engineering Journal}, \textbf{59}(6) (2020), 4843--4850. 

\bibitem{weber_1998} C. E. Weber, Pareto and the Wicksell-Cobb-Douglas functional form, \textit{Journal of the History of Economic Thought}, \textbf{20}(2) (1998), 203--210. 
   

\end{thebibliography}
\end{document}